\documentclass[twocolumn]{aastex701}

\usepackage[version=4]{mhchem}
\usepackage{amsmath,amssymb,amsfonts}%
\usepackage[T1]{fontenc}

\graphicspath{{./}{figures/}}

\begin{document}

\title{Cosmic cascades: How disk substructure regulates the flow of water to inner planetary systems}

\author[0000-0002-3291-6887]{Sebastiaan Krijt}
\altaffiliation{Authors contributed equally to this work.}
\affiliation{Department of Physics and Astronomy, University of Exeter, Exeter, EX4 4QL, UK}
\email[show]{s.krijt@exeter.ac.uk}  

\author[0000-0003-4335-0900]{Andrea Banzatti} 
\altaffiliation{Authors contributed equally to this work.}
\affiliation{Department of Physics, Texas State University, 749 N Comanche Street, San Marcos, TX 78666, USA}
\email{banzatti@txstate.edu}

\author[0000-0002-0661-7517]{Ke Zhang}
\affiliation{Department of Astronomy, University of Wisconsin-Madison, Madison, WI 53706, USA}
\email{ke.zhang@wisc.edu}

\author[0000-0001-8764-1780]{Paola Pinilla}
\affiliation{Mullard Space Science Laboratory, University College London, Holmbury St Mary, Dorking, Surrey RH5 6NT, UK}
\email{p.pinilla@ucl.ac.uk}

\author[0000-0001-8240-978X]{Till Kaeufer}
\affiliation{Department of Physics and Astronomy, University of Exeter, Exeter, EX4 4QL, UK}
\email{t.kaeufer@exeter.ac.uk}

\author[0000-0003-4179-6394]{Edwin A. Bergin}
\affiliation{Department of Astronomy, University of Michigan, 1085 S. University, Ann Arbor, MI 48109, USA}
\email{ebergin@umich.edu}

\author[0000-0003-3682-6632]{Colette Salyk}
\affiliation{Vassar College, 124 Raymond Avenue, Poughkeepsie, NY 12604, USA}
\email{cosalyk@vassar.edu}

\author[0000-0001-7552-1562]{Klaus Pontoppidan}
\affiliation{Jet Propulsion Laboratory, California Institute of Technology, 4800 Oak Grove Drive, Pasadena, CA 91109, USA}
\email{klaus.m.pontoppidan@jpl.nasa.gov}

\author[0000-0003-0787-1610]{Geoffrey A. Blake}
\affiliation{Division of Geological and Planetary Sciences, California Institute of Technology, MC 150-21, Pasadena, CA 91125, USA}
\email{gab@caltech.edu}

\author[0000-0002-7607-719X]{Feng Long}
\affiliation{Lunar and Planetary Laboratory, University of Arizona, Tucson, AZ 85721, USA}
\affiliation{NASA Hubble Fellowship Program Sagan Fellow}
\email{fenglong@arizona.edu}

\author[0000-0001-6947-6072]{Jane Huang}
\affiliation{Department of Astronomy, Columbia University, 538 W. 120th Street, Pupin Hall, New York, NY 10027, USA}
\email{jane.huang@columbia.edu}

\author[0000-0002-5296-6232]{Mar\'{i}a Jos\'{e} Colmenares}
\affiliation{Department of Astronomy, University of Michigan, Ann Arbor, MI 48109, USA}
\email{mjcolmen@umich.edu}

\author[0009-0008-8176-1974]{Joe Williams}
\affiliation{Department of Physics and Astronomy, University of Exeter, Exeter, EX4 4QL, UK}
\email{jw1436@exeter.ac.uk}

\author[0000-0001-8790-9011]{Adrien Houge}
\affiliation{Center for Star and Planet Formation, Globe Institute, {\O}ster Voldgade 5, 1350 Copenhagen, Denmark}
\email{adrien.houge@sund.ku.dk}

\author[0000-0002-0554-1151]{Mayank Narang}
\affiliation{Jet Propulsion Laboratory, California Institute of Technology, 4800 Oak Grove Drive, Pasadena, CA 91109, USA}
\email{mayank.narang@jpl.nasa.gov}

\author[0000-0002-4147-3846]{Miguel Vioque}
\affiliation{European Southern Observatory, Karl-Schwarzschild-Str. 2, 85748 Garching bei M\"{u}nchen, Germany}
\email{miguel.vioque@eso.org}

\author[0000-0001-9321-5198]{Michiel Lambrechts}
\affiliation{Center for Star and Planet Formation, Globe Institute, {\O}ster Voldgade 5, 1350 Copenhagen, Denmark}
\email{michiel.lambrechts@sund.ku.dk}

\author[0000-0003-2076-8001]{L. Ilsedore Cleeves}
\affiliation{Astronomy Department, University of Virginia, Charlottesville, VA 22904, USA}
\email{lic3f@virginia.edu}

\author[0000-0001-8798-1347]{Karin \"{O}berg}
\affiliation{Center for Astrophysics -- Harvard \& Smithsonian, 60 Garden St., Cambridge, MA 02138, USA}
\email{koberg@cfa.harvard.edu}


\collaboration{all}{and the JDISCS collaboration}

\begin{abstract}
The influx of icy pebbles to the inner regions of protoplanetary disks constitutes a fundamental ingredient in most planet formation theories. The observational determination of the magnitude of this pebble flux and its dependence on disk substructure (disk gaps as pebble traps) would be a significant step forward. In this work we analyze a sample of 21 T Tauri disks (with ages $\approx 0.5{-}2\mathrm{~Myr}$) using JWST/MIRI spectra homogeneously reduced with the JDISCS pipeline and high-angular-resolution ALMA continuum data. We find that the 1500/6000~K water line flux ratio measured with JWST - a tracer of cold water vapor and pebble drift near the snowline - correlates with the radial location of the innermost dust gap in ALMA continuum observations (ranging from 8.7 to 69~au), confirming predictions from recent models that study connections between the inner and outer disk reservoirs. We develop a population synthesis exploration of pebble drift in gapped disks and find a good match to the observed trend for early and relatively effective gaps, while scenarios where pebble drift happens quickly, gaps are very leaky, or where gaps form late are disfavored on a population level. Inferred snowline pebble mass fluxes (ranging between $10^{-6}$ and $10^{-3}~M_\oplus/\mathrm{yr}$ depending on gap position) are comparable to fluxes used in pebble accretion studies and those proposed for the inner Solar System, while system-to-system variations suggest differences in the emerging planetary system architectures and water budgets.
\end{abstract}



\section{Introduction}  \label{sec: intro}
The journey of water from molecular clouds to protoplanetary disks and ultimately planets and potential biospheres is long and complex \citep[e.g.][]{vandishoeck_2014, oberg_bergin_2021}. A critical step that occurs during the protoplanetary disk phase is the transport of water ice to the terrestrial planet formation zone through aerodynamic drift of solids \citep{whipple1972} - a fundamental process long proposed for the Solar System \citep{morfill84,stevenson88,cyr98} that could eventually be linked to the formation of potentially habitable planets \citep{krijt_2023, lichtenberg_2023}.

Seminal results on water transport in disks were achieved with the Infrared Spectrograph on the Spitzer Space Telescope \citep{irs}, by finding (anti-)correlations between \ce{H2O} emission and the disk dust mass or size, suggestive of trapping of water ice beyond the snowline in massive or large disks \citep{najita_2013,banzatti_2020}. The MIRI spectrograph on JWST \citep{miri,miri2,miri3} now probes physical and chemical conditions in the inner few au of planet-forming disks with unprecedented combination of high spectral resolution and sensitivity at infrared wavelengths. Analysis of the first MIRI-MRS disk spectra in Cycle 1 revealed a rich chemical diversity ranging from water-dominated to hydrocarbon-rich systems \citep{tabone_2023, grant23, banzatti_2023, kamp2023, pontoppidan_2024, henning2024, arulanantham2025}. The sharper spectral view of MIRI has already improved upon previous Spitzer results by revealing that trends with dust disk radii are due to a cold ($\sim$170--400$\mathrm{~K}$) water vapor reservoir that is enhanced in compact disks that are considered to be drift-dominated \citep{banzatti_2023, romero-mirza_2024, banzatti_2025}, supporting early predictions of ice delivery to the snowline by pebble drift \citep{cieslacuzzi2006}. However, recent modeling works show that more than traced by the outer disk radius, pebble drift into the snowline region should be regulated by the radial location and depth of disk gaps acting as pebble traps \citep{kalyaan_2021, kalyaan_2023, easterwood_2024, mah_2024}. A recent analysis of 8 compact disks of intermediate sizes (40--60~au) indeed finds a range in cold water enrichment that does not simply depend on the outer disk size \citep{temmink_2025}\footnote{This work included only one disk more compact than $\sim 40$~au, therefore focusing on intermediate-size disks.}, supporting discussions of larger samples in \citet{banzatti_2025,gasman2025}.

In the past 10 years, ALMA has indeed revealed that the radial distributions of dust and pebbles in disks are often not smooth but show sub-structures in the form of rings and gaps \citep{andrews2018,andrews_2020}, which can slow down and reduce pebble and water ice delivery to the inner disk \citep{kalyaan_2023, mah_2024}. While models are quickly progressing and finding some support for the ice drift scenario to explain the correlations found with infrared water spectra, some aspects of the problem are emerging as key to this regulatory process. One aspect is the degree of `leakiness' of disk gaps as pebble traps \citep{stammler_2023, pinilla_2024, tong_2025, kalyaan_2023, easterwood_2024, huang_2025}, which should be directly linked to the amount of inner disk water enrichment by drift (with larger enrichment provided by leakier traps). A second major aspect is the timing of gap opening in the disk, which should typically start before the age of protoplanetary disks observed in nearby star-forming regions ${\sim}1{-}5$~Myr \citep{segura-cox2020,ALMA_HLTau,sheehan2020}. A third aspect is the range of masses and radii disks may have when they form, which should provide different initial conditions of ice reservoirs for pebble drift. All these aspects could play a role in producing the scatter in inner disk water enrichment recently observed in larger samples with MIRI \citep[see discussions in][]{banzatti_2025, gasman2025, arulanantham2025}.

\begin{figure*}[ht!]
\includegraphics[width=\textwidth]{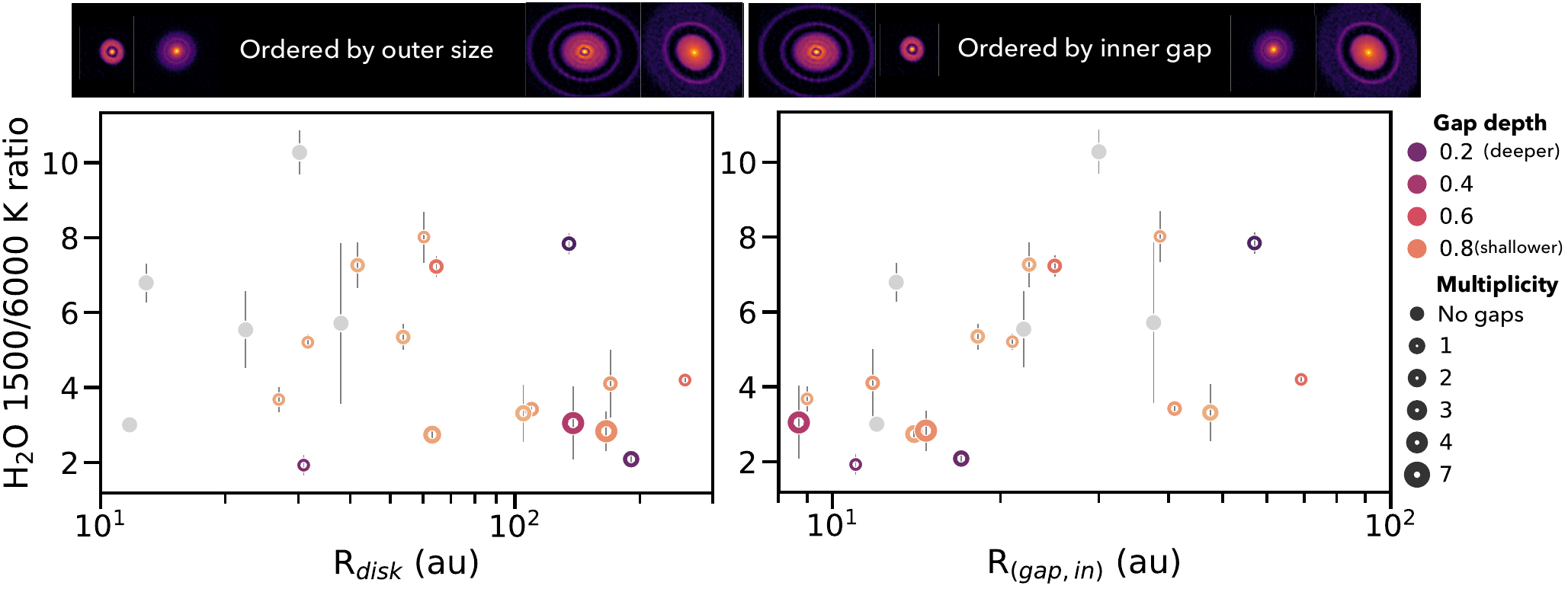}
\caption{Trends observed between the cold water vapor reservoir in the inner disk (traced by the 1500/6000 K line flux ratio) and the dust disk radius (left) or inner disk gap location (right). The data is color-coded to illustrate the gap depth (darker color indicates a deeper gap) and the symbol size indicates the number of gaps detected in the disk, including at larger radii. Five disks without detected gaps at the current spatial resolution are also included for comparison in grey.} 
\label{fig:data_trends}\end{figure*}

\section{Data and analysis}\label{sec:data}
This analysis requires using high quality data from MIRI and high spatial resolution ALMA observations to resolve dust gaps in inner disks. For MIRI, we adopt the samples published in \cite{banzatti_2025,gasman2025} all homogeneously reduced with the JDISCS pipeline, which adopts the standard MRS pipeline \citep{MIRI_pip} up to stage 2b and then uses asteroid calibrators to obtain the highest quality fringe correction, achieving signal-to-noise (S/N) of 200--400 \citep{pontoppidan_2024}. As in \cite{banzatti_2025}, we used the JDISCS reduction version 8.2 that used the MRS pipeline version 11.17.19 and Calibration Reference Data System context jwst\_1253.pmap. The JWST data come from the following programs from the JDISCS and MINDS collaborations: GO-1282 (PI: T. Henning; co-PI: I. Kamp), GO-1549 (PI: K. Pontoppidan), GO-1584 (PI: C. Salyk; co-PI: K. Pontoppidan), GO-1640 (PI: A. Banzatti). For the ALMA data, we limit to high-resolution images as obtained in the DSHARP survey \citep{andrews2018} and the \citet{long_2019} survey of resolved disks in Taurus, which achieved spatial resolutions better than ${\sim}0''.1$ (${\approx}15 \mathrm{~au}$ at 150 pc). This work includes only T Tauri stars, and the specific sample available from the combined ALMA and JWST data covers stellar masses of 0.2--1~M$_{\odot}$ and ages of ${\sim}0.5{-}2\mathrm{~Myr}$ (see below). To have a sample that should focus on pebble drift effects, we also exclude disks where other processes are known to affect the observed water spectrum. Specifically, we exclude disks that have a spatially-resolved millimeter dust cavity \citep[which is known to deplete the hot water reservoir, see][and Mallaney et al. in prep.]{salyk_2015,banzatti_2017}, disks in a wide stellar binary system \citep[which is known to change pebble drift efficiency in the disk,][]{zagaria_2021}, systems known to be episodic accretion bursters \citep[which have been found to temporarily increase the cold water reservoir,][]{smith_2025}, and disks that are viewed at an inclination angle higher than 65 deg (which causes inner disk obscuration and makes it hard to resolve inner gaps).
With these criteria, we collected from published work a sample of 21 disks with properties reported in Table \ref{tab: sample}.

To enable a consistent comparison of relative ages across our sample, we compiled effective temperatures and stellar luminosities from \citet{andrews2018,Manara_PPVII} and estimate stellar masses and ages by comparing them to evolutionary tracks. For this, we used the \texttt{Python} package \texttt{ysoisochrone}\footnote{\url{https://github.com/DingshanDeng/ysoisochrone}} \citep{Deng_2025_ysoisochrone}, which uses evolutionary tracks of \citet{Feiden16} for sources with $T_{\rm eff}>3900\mathrm{~K}$ and that of \citet{Baraffe15} for $T_{\rm eff} \le 3900\mathrm{~K}$. The typical uncertainty in the ages is ${\sim}0.5~\mathrm{Myr}$ and in the majority of cases, the age estimates are in good agreement with values reported in the literature \citep[e.g.,][]{andrews2018, long_2019, romero-mirza_2024}. In a few instances, however, estimates from \texttt{ysoisochrone} are younger. We adopt our homogeneous approach to ensure internal consistency across the dataset. Lastly, we list in Table~\ref{tab: sample} the approximate spatial resolution achieved by ALMA, typically around ${\sim}4{-}5~\mathrm{au}$ for the DSHARP sources and ${\sim}8{-}9\mathrm{~au}$ for other objects in Taurus with visibility analyses. The majority of sources thus have resolutions within a factor of $2{-}3$ and fall in the `high' and `moderate' angular resolution scenarios of \citet[][Sect.~3.2]{mah_2024}. We remark that even with the state-of-the-art resolution and analyses achieved in previous ALMA studies, in some cases the gap depth may be underestimated and (very) close-in and/or shallow gaps may have been missed (see Section ~\ref{sec:discussion}).

Figure \ref{fig:data_trends} shows the cold water enrichment diagnostic, the 1500/6000~K flux ratio of water lines observed with MIRI as defined in \cite{banzatti_2025}. Essentially, this line flux ratio measures the strength of lower-energy (upper level energy of $\sim1500$~K) water transitions in comparison to higher-energy (upper level energy of $\sim6000$~K) transitions and has been found to reflect the observed cold water mass in the disk surface \citep[][and Appendix \ref{sec:appendix_B}]{romero-mirza_2024,banzatti_2025}. The horizontal axes in the figure are the dust disk radius including 90--95\% of the millimeter flux observed with ALMA (left) and the innermost disk gap location R$_{gap, in}$ (right). Several disks have additional gaps at larger radii (see Table \ref{tab: sample}), but we only use the innermost one as models predict it should have the strongest effect on water delivery (Section \ref{sec: intro}).
For comparison to the gapped disks, we also include disks that have no gaps detected at the same spatial resolution; for these disks (five in total) as R$_{gap, in}$ in the right panel of Figure \ref{fig:data_trends} we use the dust disk size itself, since no inner gaps are detected inward of that. 

While previous work found a significant anti-correlation with dust disk radius \citep{banzatti_2020,banzatti_2023,romero-mirza_2024}, interpreted as a proxy for pebble drift efficiency over the disk as a whole, the larger sample included in this work is only weakly correlated (Pearson correlation coefficient of $-0.2$ and p-value of 0.47), as already discussed for similarly large samples in \cite{banzatti_2025,gasman2025}. Using the same exact sample, the water diagnostic ratio is instead significantly correlated with the inner gap location (Pearson correlation coefficient of 0.5 and p-value of 0.06), directly confirming earlier model predictions \citep{kalyaan_2023}. It is important to remark that models do not simply predict a linear correlation between the two, due to the age spread and the steep decline of the water enrichment curve in case of outer disk gaps \citep[e.g. Fig.~3 in][]{kalyaan_2023}. Additionally, variations in inner disk temperature structure and morphology (e.g., flaring) may increase the spread in diagnostic ratios. In fact, the data in Figure \ref{fig:data_trends} show a large spread of datapoints for gaps at 30--60~au, which is consistent with the spread in cold/hot water ratio measured in compact disks that have this range of sizes \citep{temmink_2025}. It is also worth noting that some disks with deep inner gaps (e.g., CI Tau) also have large dust disk sizes due to the presence of multiple gaps. Together with smooth compact disks, these large disks with inner deep gaps were driving the previously reported anti-correlation between inner water enrichment and disk size in smaller samples \citep{banzatti_2023,banzatti_2025}.

\begin{deluxetable}{l c c |  c c c c c l l | c c c c c}
\tabletypesize{\footnotesize}
\tablewidth{0pt}
\tablecaption{\label{tab: sample} Sample properties for disks included in this work.}
\tablehead{ Name & JWST PID & \ce{H2O} ratio & $R_\mathrm{gap}/\mathrm{au}$ & $W_\mathrm{gap}/\mathrm{au}$ & $D_\mathrm{gap}$ & \# gaps & $\log R_\mathrm{disk}/\mathrm{au}$ & Res/au & Ref. & age/Myr & $M_*/M_\odot$} 
\tablecolumns{8}
\startdata
        AS 209 & 1584 & 3.05 & 8.7 & 4.7 & 0.45 & 7 & 2.14 & 4.4 & a & 0.93 & 0.80  \\ 
        BP Tau & 1282 & 7.27 & 22.5 & 9.3 & 0.99 & 2 & 1.62 & 7.1$^\mathrm{v}$ & b,c & 0.56 & 0.54 \\ 
        CI Tau & 1640 & 2.09 & 17 & 9.2 & 0.16 & 3 & 2.28 & 7.9 & d & 0.62 & 0.70 \\ 
        CX Tau & 1282 & 5.71 & \nodata & \nodata & \nodata & 0 & 1.58 & 4.1 & e  & 1.00 &0.34 \\ 
        DL Tau & 1282 & 2.83 & 14.7 & 4.2 & 0.86 & 7 & 2.22 & 8.7$^\mathrm{v}$ & b,c & 0.67 &0.70 & \\ 
        DoAr 33 & 1584 & 3.68 & 9 & 1 & 0.95 & 1 & 1.43 & 3.4 & a & 1.48 & 1.01 \\ 
        DR Tau & 1282 & 5.35 & 18.2 & 1.7 & 0.99 & 2 & 1.73 & 9.8$^\mathrm{v}$ & b,c &  0.54 & 0.68 & \\ 
        Elias 20 & 1584 & 7.23 & 25 & 3.5 & 0.75 & 2 & 1.81 & 3.2 & a & 0.50 & 0.52 \\ 
        Elias 24 & 1584 & 7.84 & 57 & 22.8 & 0.03 & 2 & 2.13 & 4.7 & a & 0.50 & 0.71 \\ 
        Elias 27 & 1584 & 4.20 & 69.1 & 14.3 & 0.73 & 1 & 2.41 & 5.5 & a &  0.74 & 0.58 \\ 
        FZ Tau & 1549 & 3.00 & \nodata & \nodata & \nodata & 0 & 1.07 & 5.7 & f & 0.50 & 0.48  \\ 
        GK Tau & 1640 & 6.79 & \nodata & \nodata & \nodata & 0 & 1.11 & 7.1$^\mathrm{v}$ & b,c &  0.62 & 0.64 \\ 
        GO Tau & 1640 & 4.10 & 11.8 & 2.7 & 0.91 & 2 & 2.23  & 7.9$^\mathrm{v}$ & b,c & 1.58 & 0.35 \\ 
        GQ Lup & 1640 & 10.3 & \nodata & \nodata & \nodata & 0 & 1.75 & 4.7 & g,h & 0.60 & 0.67\\ 
        GW Lup & 1282 & 3.31 & 47.5 & 2.4 & 0.99 & 3 & 2.02 & 6.6 & a & 1.62 & 0.42 \\ 
        HP Tau & 1640 & 5.54 & \nodata & \nodata & \nodata & 0 & 1.35 &  9.7$^\mathrm{v}$ & b,c & 0.81 & 0.88 \\ 
        IQ Tau & 1640 & 3.42 & 41 & 5.6 & 0.92 & 2 & 2.04 & 7.2$^\mathrm{v}$ & b,c & 0.56 & 0.48 \\ 
        RU Lup & 1584 & 2.74 & 14 & 1 & 0.95 & 4 & 1.8  & 3.9 & a & 0.63 & 0.64\\ 
        SR 4 & 1584 & 1.93 & 11 & 6.3 & 0.23 & 1 & 1.49 & 4.6 & a &  0.77 & 0.65 \\ 
        Sz 114 & 1584 & 8.02 & 38.6 & 4.3 & 0.94 & 1 & 1.78 & 4.6 & a &  1.04 & 0.17 \\ 
        WSB 52 & 1584 & 5.21 & 21 & 1 & 0.95 & 1 & 1.5 & 3.7 & a& 0.67 & 0.47 \\ 
\enddata
\tablecomments{
The \ce{H2O} ratio is the 1500/6000~K line flux ratio that is sensitive to the cold water mass relative to the hot water mass in each disk, as defined in \cite{banzatti_2025}. $R_\mathrm{gap}$, $W_\mathrm{gap}$, $D_\mathrm{gap}$, and \# gaps are the inner gap radius, width, depth, and total number of gaps detected in each disk as reported in \cite{huang_2018,zhang_2023,gasman2025}. For disks without gap detections, we only use the outer radius that includes 90--95\% of the millimeter emission ($R_{\rm{disk}}$) as reported in \cite{huang_2018,long_2019,facchini_2019,Guerra-Alvarado_2025}. Spatial resolution (Res) is estimated as the distance $\times$ minor beam size (for data analyzed in the image plane), or (1/2) $\times$ distance $\times$ minor beam size (for data analyzed in the visibility plane, indicated by superscript `$^\mathrm{v}$'). ALMA references: a = \citet{huang_2018}, b = \citet{zhang_2023}, c = \citet{long_2019}, d = \citet{clarke2018}, e = \citet{facchini_2019}, f = Long et al. (in prep.), g = \citet{wu2017}, h = \citet{Guerra-Alvarado_2025}. Stellar ages and masses are obtained as explained in Sect.~\ref{sec:data}.}
\end{deluxetable}

\section{Model exploration}\label{sec:models}

In order to test the observed trends presented in the previous section, we perform a population synthesis study that reproduce the outcome of dust evolution and pebble drift in disks of similar conditions to the observed sample, with a series of modeling steps as follows.

\subsection{Estimating pebble mass fluxes in disks with gaps}\label{sec:pebble_models}

The temporal evolution of the pebble flux arriving at the inner disk is modeled using \texttt{pebble predictor}\footnote{\url{https://github.com/astrojoanna/pebble-predictor}} \citep{drazkowska_2021}. The gas surface density is described as a tapered power-law (characterized by a size $R_\mathrm{taper}$) and pebble drift is calculated self-consistently assuming a combination of drift-limited and fragmentation-limited dust coagulation processes \citep[see e.g.,][]{birnstiel_2024}. \texttt{pebble predictor} has been successfully benchmarked against \texttt{dustpy} \citep{stammler_2022} for a range of disk masses, turbulence levels, and dust fragmentation velocities \citep[][Fig. 7]{drazkowska_2021}. Pebble fluxes are calculated at the midplane water snowline (defined as the position where $T_\mathrm{disk}=150\mathrm{~K}$) 
and we follow the approach of \citet[][Table~1]{mulders_2021} to update the disk temperature profile when modeling different stellar masses. We use relatively fragile grains with constant fragmentation velocity $v_\mathrm{f} = 3\mathrm{~m/s}$, in line with recent laboratory, observational, and modeling studies favoring relatively fragile grains \citep{musiolik2019,jiang_2024,ueda2024,williams2025}.

\begin{figure*}[ht!]
\centering
\includegraphics[width=0.95\textwidth]{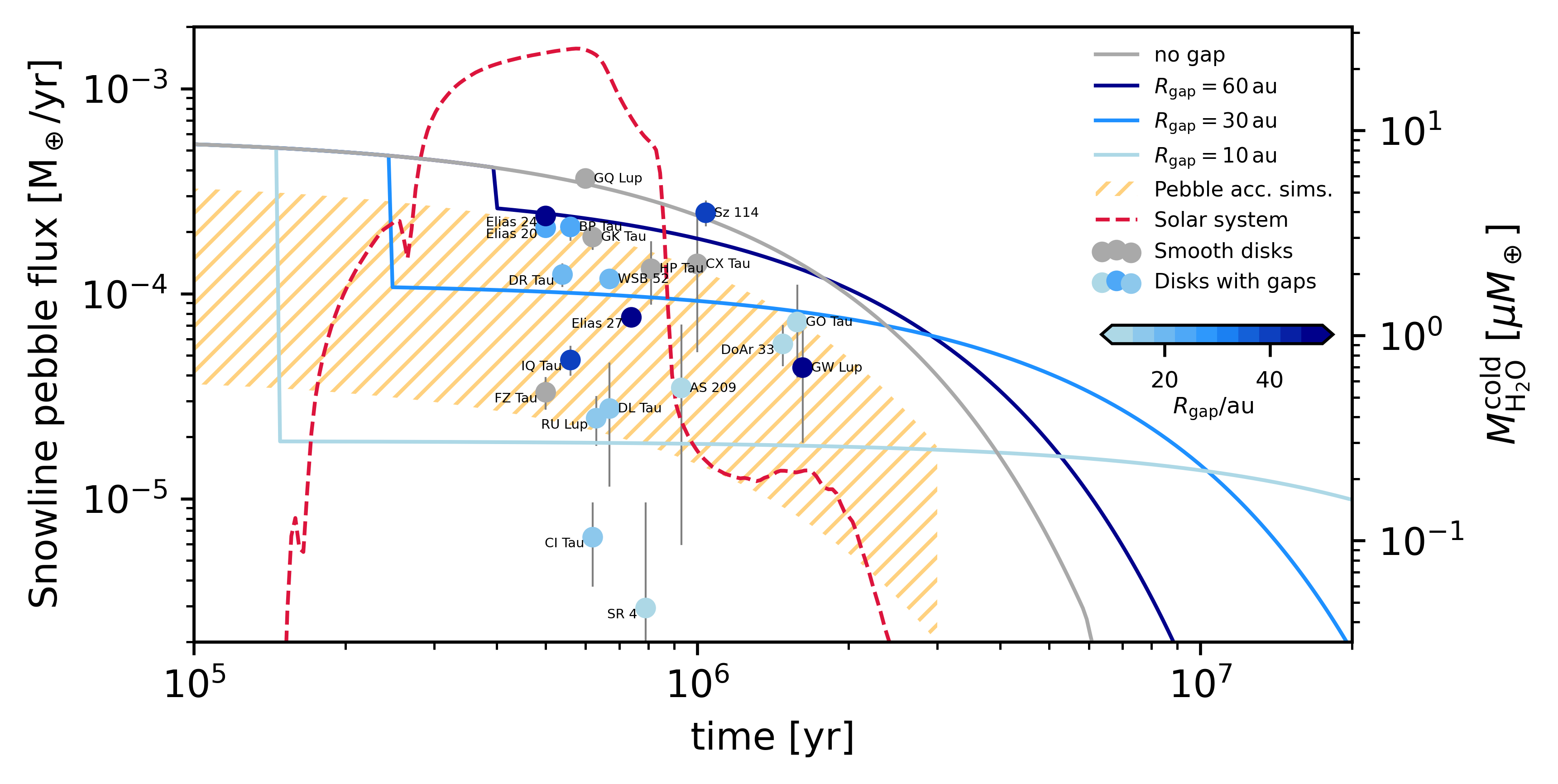}
\caption{Compilation of pebble mass flux vs. time evolutions. Solid lines show models developed in this study for various gap positions (see Sect.~\ref{sec:pebble_models}), which can be compared to mass fluxes required in pebble accretion simulations to form diverse inner planetary systems (dashed area, see \citealt{lambrechts_2019}) and the pebble mass flux inferred for the inner solar system using models capable of matching the NC/CC isotopic dichotomy (red dashed, see \citealt{lichtenberg_2021}). Symbols show ages and cold water masses of all 21 objects in our sample computed from JWST/MIRI line ratios (Table~\ref{tab: sample}) in combination with Eq.~\ref{eq:h2oratio} (see Appendix~\ref{sec:appendix_B}), and the left and right y-axes are related through Eq.~\ref{eq:carlos}. Symbol color corresponds to inner gap position as obtained from high-spatial-resolution ALMA observations (see Sect.~\ref{sec:data}) and error-bars reflect uncertainties in measured line ratios.}\label{fig:pebble_flux_overview}
\end{figure*}

We expand these models to include one gap per disk, following previous work that showed the innermost gap, if deep enough, will dominate the regulatory process of water delivery by pebble drift even in the case of multiple gaps \citep{kalyaan_2023,easterwood_2024}. In our models, gaps are described by three parameters: their location $R_\mathrm{gap}$, the time at which they appear $t_\mathrm{gap}$, and an effective `impact' $\Lambda_\mathrm{gap}$ on the mass flux arriving at the water snowline -- a parameter that in principle can depend on gap depth, width, but also its position and dust/pebble properties.

To synthesize lessons from dedicated studies investigating the impact of disk substructure on local pebble flux rates and inner disk water enrichment \citep{kalyaan_2021, kalyaan_2023, stammler_2023, easterwood_2024, mah_2024, tong_2025, huang_2025}, we implement the appearance of outer disk substructure in \texttt{pebble predictor} as follows: once a gap opens (at position $R_\mathrm{gap}$ and time $t_\mathrm{gap}$), its impact on the snowline pebble flux is not felt until a time
\begin{equation}\label{eq:delay}
t^\mathrm{SL}_\mathrm{gap} = t_\mathrm{gap}+t_\mathrm{delay} = t_\mathrm{gap} + \frac{R_\mathrm{gap} - R_\mathrm{SL}}{\tilde{v}_\mathrm{peb}},
\end{equation}
where $R_\mathrm{SL}$ is the water snowline position and   $\tilde{v}_\mathrm{peb}$ is a representative pebble drift speed that we set to $\tilde{v}_\mathrm{peb} = 2\times10^{-4}\mathrm{~au/yr}$ throughout\footnote{Pebble drift velocities depend on grain size and various disk parameters, with maximum values typically around $50 \mathrm{~m/s} \simeq 10^{-2} \mathrm{~au/yr}$ \citep{birnstiel_2024}, however, even the largest particles in the outer disk will generally have Stokes numbers $\mathrm{St}\ll1$. Our choice of a velocity of  $2\times10^{-4}\mathrm{~au/yr}$ matches well the observed delay in inner disk mass reduction shown in the more comprehensive models of \citet{kalyaan_2023}.}. At times $t>t^\mathrm{SL}_\mathrm{gap}$, we multiply the pebble flux through the disk by a factor $\Lambda_\mathrm{gap}(<1)$ to mimic the gap's ability to prevent material from drifting all the way to the snowline unhindered. We set
\begin{equation}\label{eq:gap_impact}
\Lambda_\mathrm{gap} =  1 - f^\mathrm{local}_\mathrm{gap} \times \exp\left[-\left(\frac{R_\mathrm{gap}}{R_\mathrm{taper}} \right)^\gamma\right]
\end{equation}
with $f^\mathrm{local}_\mathrm{gap}$ the fractional pebble flux that the gap blocks \emph{locally} at $R_\mathrm{gap}$. With the focus being on relatively deep gaps, we use $f^\mathrm{local}_\mathrm{gap} = 0.99$ but vary this later. The exponential term in Eq.~\ref{eq:gap_impact} acts to reduce the impact far-out gaps have on the inner disk, as even though these gaps may be efficient filters locally, only a small fraction of the disk ice mass will be located exterior to their position \citep[e.g.,][]{easterwood_2024,williams2025}. We set $\gamma=2$, for which we find the reduction in mass reaching the inner disk matches well the behavior in \citet[][Fig.~3]{kalyaan_2023}.

The blue curves in Fig.~\ref{fig:pebble_flux_overview} show four representative models ($M_*=M_\odot$, $M_\mathrm{disk}=0.1M_*$, $\alpha_\mathrm{turb}=3\times10^{-4}$, $v_\mathrm{frag}=3\mathrm{~m/s}$, $R_\mathrm{taper}=60\mathrm{~au}$). While all three models with gaps use $t_\mathrm{gap}=10^5\mathrm{~yr}$ (and identical $f^\mathrm{local}_\mathrm{gap}$), closer-in gaps can be seen to lead to an earlier and more dramatic reduction in the pebble flux at the water snowline, ultimately allowing for a longer-lived (but lower) influx of material to the inner disk. The results of our semi-analytic model then broadly capture the relevant behavior observed in detailed simulations of \citet{kalyaan_2021,kalyaan_2023, easterwood_2024, mah_2024}, while offering flexibility to explore various scenarios (see~Sect.\ref{sec:parameter_exploration}).

\subsection{Pebble drift, cold water, and 1500/6000K ratios}\label{sec:drift_to_cold_water}
Next, we use the approach detailed in \citet{romero-mirza_2024} to connect the pebble mass fluxes crossing the midplane water snowline $F^\mathrm{SL}_\mathrm{peb}(t)$ to an observable cold water mass $M^\mathrm{cold}_{\ce{H2O}}(t)$ in the disk surface layers overhead:
\begin{equation}\label{eq:carlos}
F^\mathrm{SL}_\mathrm{peb}(t) = \frac{M^\mathrm{cold}_{\ce{H2O}}(t)}{t_{\ce{H2O}}} \frac{\eta}{f_\mathrm{ice} f_{\ce{H2O}}},
\end{equation}
where $f_\mathrm{ice}$ is the mass fraction of pebbles that is ice, $f_{\ce{H2O}}$ is the mass fraction of ice that is water ice, $\eta$ is correction factor accounting for the unobservable water column beneath the optically thick layer, and $t_{\ce{H2O}}$ is the typical timescale on which newly-delivered water is removed/processed (either by chemistry or transport processes). Following \citet{romero-mirza_2024}, we set $\eta=10^3$, $f_\mathrm{ice}=0.2$, $f_{\ce{H2O}} =0.8$, and $t_{\ce{H2O}} = 100~\mathrm{yr}$ and assume these parameters do not vary with time or between disks \citep[see, however,][]{sellek2025, houge_2025, kanwar2025}. With this approach, we are effectively assuming that the primary source of the cold (${\lesssim}300{-}400\mathrm{~K}$) water vapor reservoir in the disk surface is the sublimation of icy pebbles in the obscured midplane below, with the released vapor quickly being mixed upwards where it is observed \citep[see also][]{banzatti_2023, banzatti_2025, houge_2025}. In Fig.~\ref{fig:pebble_flux_overview}, Eq.~\ref{eq:carlos} is used to connect cold water masses (right-hand y-axis) to snowline pebble mass flux (left-hand y-axis).

We directly connect cold water water masses to 1500/6000K line flux ratios using an empirical relationship based on an investigation of published model series augmented by a new suite of multi-component slab models (see Appendix \ref{sec:appendix_B}). The adopted relation takes the form
\begin{equation}\label{eq:h2oratio}
1500/6000\mathrm{K} = \mathcal{A} + \left(\frac{M^\mathrm{cold}_{\ce{H2O}}}{0.4 ~ \mathrm{\mu} M_\oplus}\right)^{q}
\end{equation}
where we set $\mathcal{A}=1.75$ and $q=0.8$ to match the behavior observed in slab models (see Fig.~\ref{fig:hockey_stick_figure}). Using Eq.~\ref{eq:h2oratio}, we can compute cold water masses (and hence pebble fluxes) for all objects in Table~\ref{tab: sample} and compare these to values from dust evolution models, pebble accretion models, and the Solar System\footnote{There is ongoing debate about the degree to which material could flow into the inner Solar System during the first few Myr \citep{kruijer_2020}. We only show one representative example in Fig.~\ref{fig:pebble_flux_overview} but note that other recent studies find comparable pebble mass fluxes and time evolutions \citep[see][]{colmenares2024}.} in Fig.~\ref{fig:pebble_flux_overview}. While the populations overlap, the majority of the smooth disks (and disks with exclusively far-away gaps) are seen to cluster towards high pebble fluxes, while disks with close-in gaps include considerably lower fluxes.

\begin{figure*}[ht!]
\centering
\includegraphics[width=0.87\textwidth]{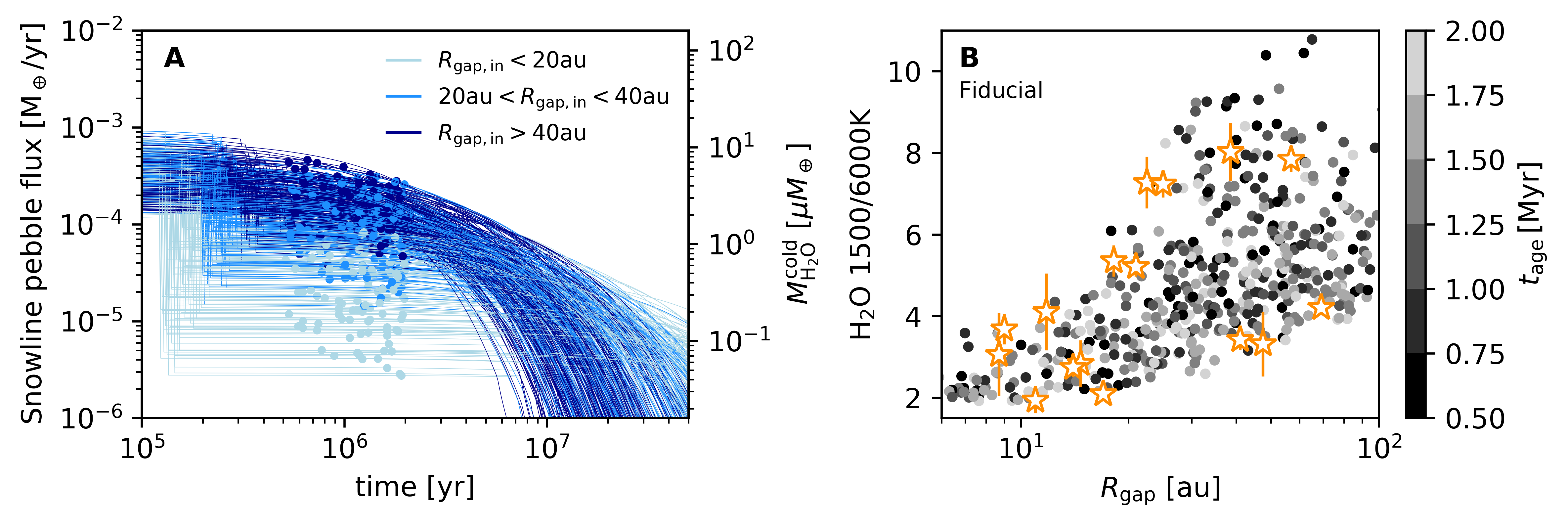}
\caption{Results for our fiducial model series described in Sect.~\ref{sec:popsynth}. \textbf{A:} Collection of pebble drift models in disks with gaps appearing at different times and positions (see Sect.~\ref{sec:popsynth}). For every model a single time between $\mathbf{0.5}<t_\mathrm{age}/\mathrm{Myr}<\mathbf{2}$ is selected at random (shown by the small dots). \textbf{B:} \ce{H2O} 1500/6000K diagnostic ratio for every model in panel A at their respective $t_\mathrm{age}$, calculated from the pebble mass flux using Eqs.~\ref{eq:carlos} and \ref{eq:h2oratio}.  Gold stars show data from Table~\ref{tab: sample}.}\label{fig:popsynth_1}
\end{figure*}

\subsection{Reproducing population-level trends}\label{sec:popsynth}
We present in Fig.~\ref{fig:popsynth_1}A a total of $N=500$ unique models using a variation of star, disk, and gap parameters. For every iteration we randomly select a stellar mass $M_\star/M_\odot = [0.4,1]$, a gas disk mass $\log_{10}(M_\mathrm{disk}/M_\star) = [-1,-0.5]$, a disk size $R_\mathrm{taper}/\mathrm{au} = [30,80]$ (sampled logarithmically), and gap opening location $R_\mathrm{gap} = [5~\mathrm{au},1.5R_\mathrm{taper}]$. For the fiducial modeling series (see Table~\ref{tab:model_params}) we set the turbulence strength to $\alpha_\mathrm{turb} = 10^{-3}$, dust-to-gas ratio to 0.02, fragmentation velocity $v_\mathrm{f}=3~\mathrm{m/s}$, and gap opening time to $t_\mathrm{gap}=10^5\mathrm{~yr}$. Initial disk masses and sizes are in line with results of star formation simulations \citep{bate_2018} and results of the AGE-PRO ALMA large program \citep{zhang2025, tabone2025}, while turbulence levels are based on a variety of empirical constraints \citep{rosotti_2023, jiang_2024}. We then sample each model exactly once at a random time between 0.5 and 2 Myr (roughly the age spread in our sample) and convert the pebble mass flux at that time into a 1500/6000~K ratio using Eqs.~\ref{eq:carlos} and \ref{eq:h2oratio}. The results of this exercise are shown in Fig.~\ref{fig:popsynth_1}B, together with the data from Table~\ref{tab: sample} (excluding the smooth disks). The models can be seen to yield results that are very consistent with the data: both avoiding the region in the top-left part of the plot (i.e., high line ratios with close-in gaps), while allowing for a range of 1500/6000~K ratios for (inner) gaps further out.

\begin{deluxetable*}{lccccl}
\tablewidth{0pt}
\tablecaption{Overview of parameters used in the disk population synthesis series of Sect.~\ref{sec:popsynth} and \ref{sec:parameter_exploration}.\label{tab:model_params}}
\tablehead{
\colhead{Series} & Fig. &\colhead{$\alpha_\mathrm{turb}$} & \colhead{$v_\mathrm{frag}$} & \colhead{$t_\mathrm{gap}/\mathrm{yr}$} & \colhead{Gap treatment}}
\tablecolumns{6}
\startdata
Fiducial & \ref{fig:popsynth_1}A,B & $10^{-3}$ & $3\mathrm{~m/s}$ & $10^5$ & Eq.~\ref{eq:gap_impact} and $f_\mathrm{gap}^\mathrm{local}=0.99$\\
Rapid drift & \ref{fig:popsynth_2}A,B & $10^{-4}$ & $9\mathrm{~m/s}$ & - & - \\
Leaky gaps & \ref{fig:popsynth_2}C,D & - & - & - & $f_\mathrm{gap}^\mathrm{local}=[0.5-0.9]$\\
Later gaps & \ref{fig:popsynth_2}E,F & - & - & $10^6$ & -\\
Indiscriminate gaps & \ref{fig:popsynth_2}G,H & - & - & - & $\Lambda_\mathrm{gap} =  (1 - f^\mathrm{local}_\mathrm{gap})$ and $f_\mathrm{gap}^\mathrm{local}=[0.5-0.99]$
\enddata
\tablecomments{Entries marked `-' default to the fiducial model. For the Later gaps case, ages are sampled only between 1-2 Myr in Fig.~\ref{fig:popsynth_2} to avoid depicting systems with gaps that have not formed yet.}
\end{deluxetable*}

\subsection{On the timing of pebble drift and "leaky" gaps}\label{sec:parameter_exploration}
\begin{figure*}[ht!]
\centering
\includegraphics[width=0.87\textwidth]{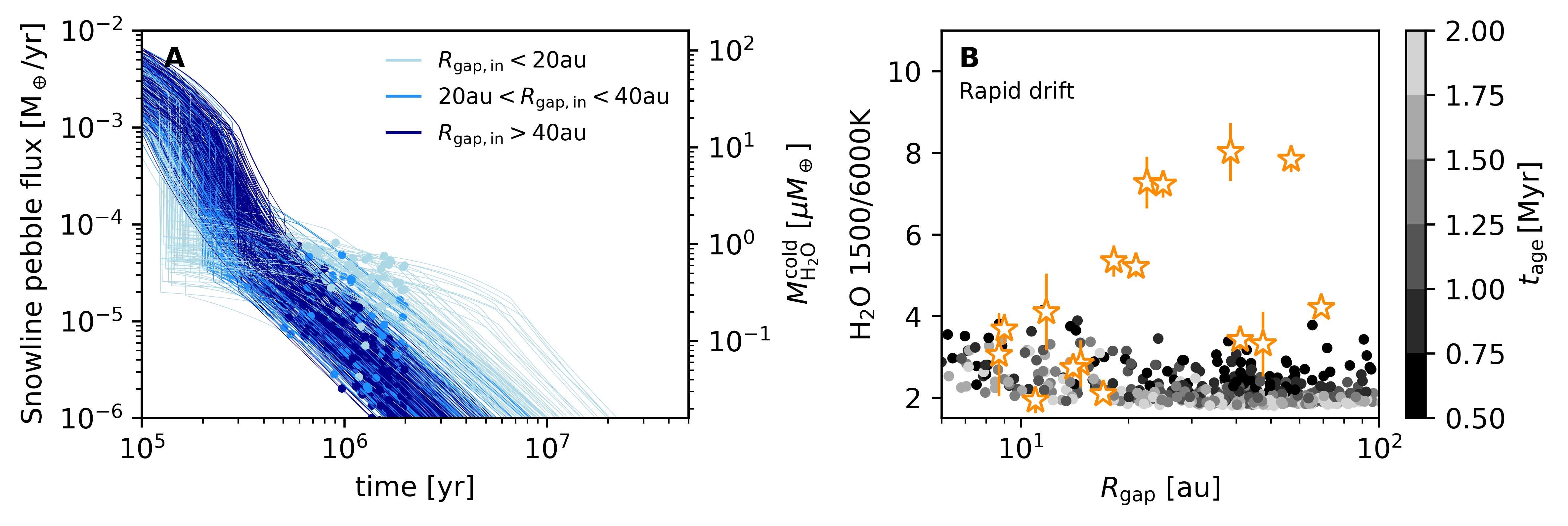}
\includegraphics[width=0.87\textwidth]{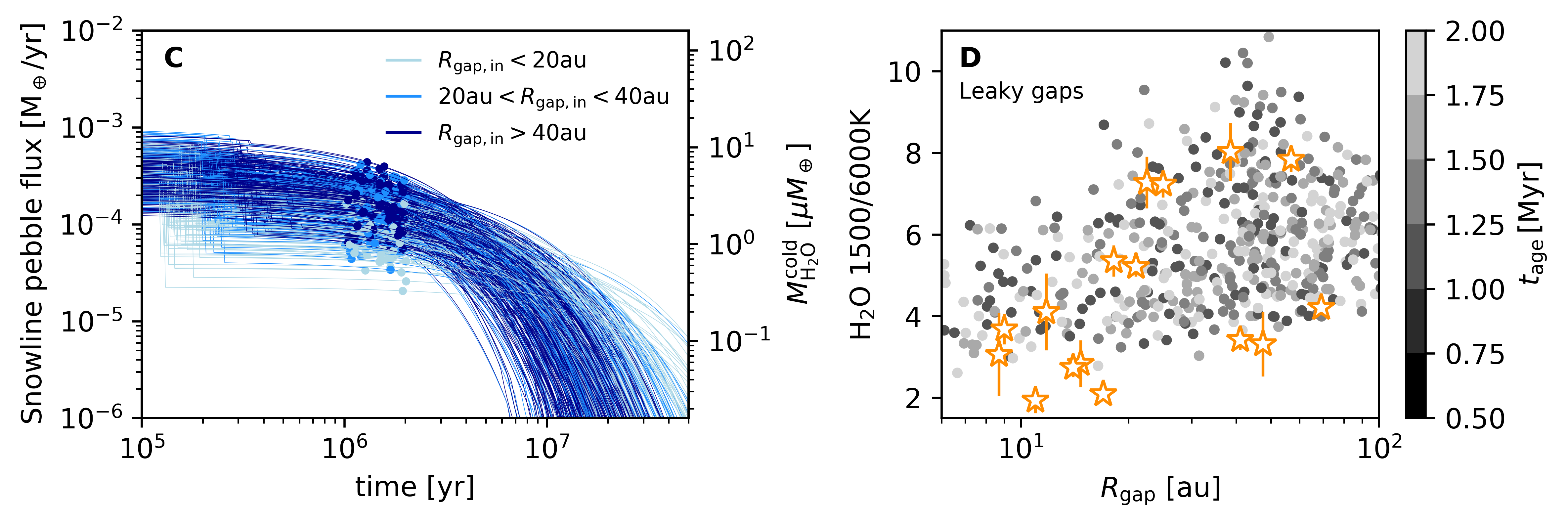}
\includegraphics[width=0.87\textwidth]{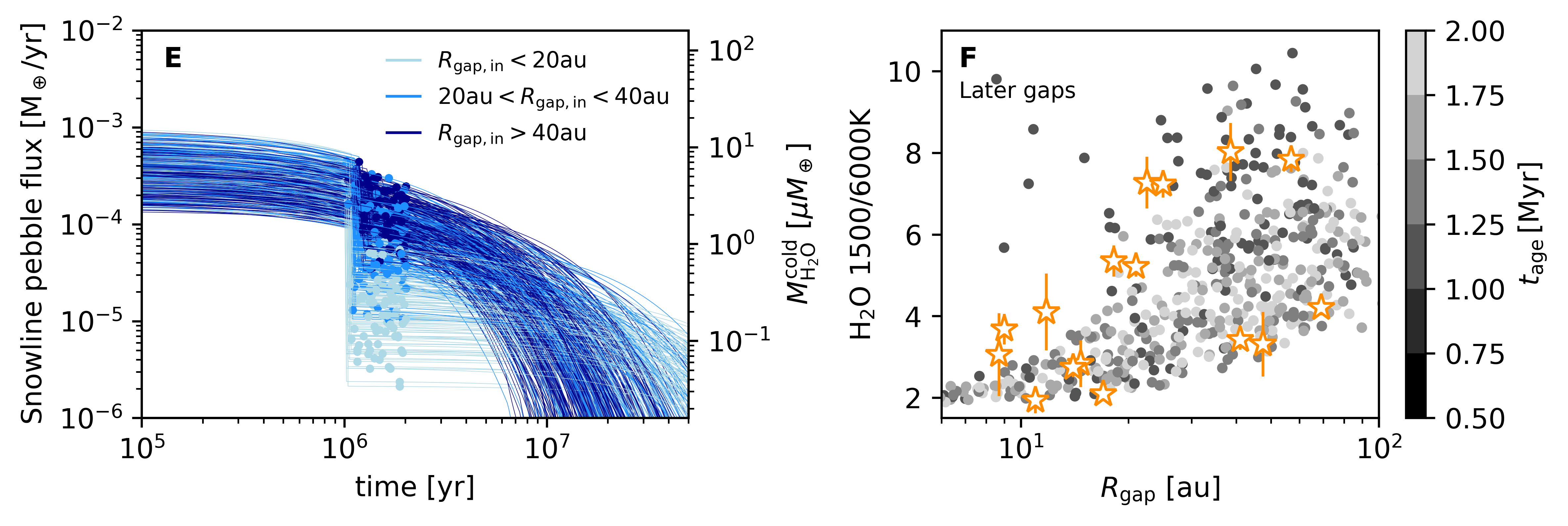}
\includegraphics[width=0.87\textwidth]{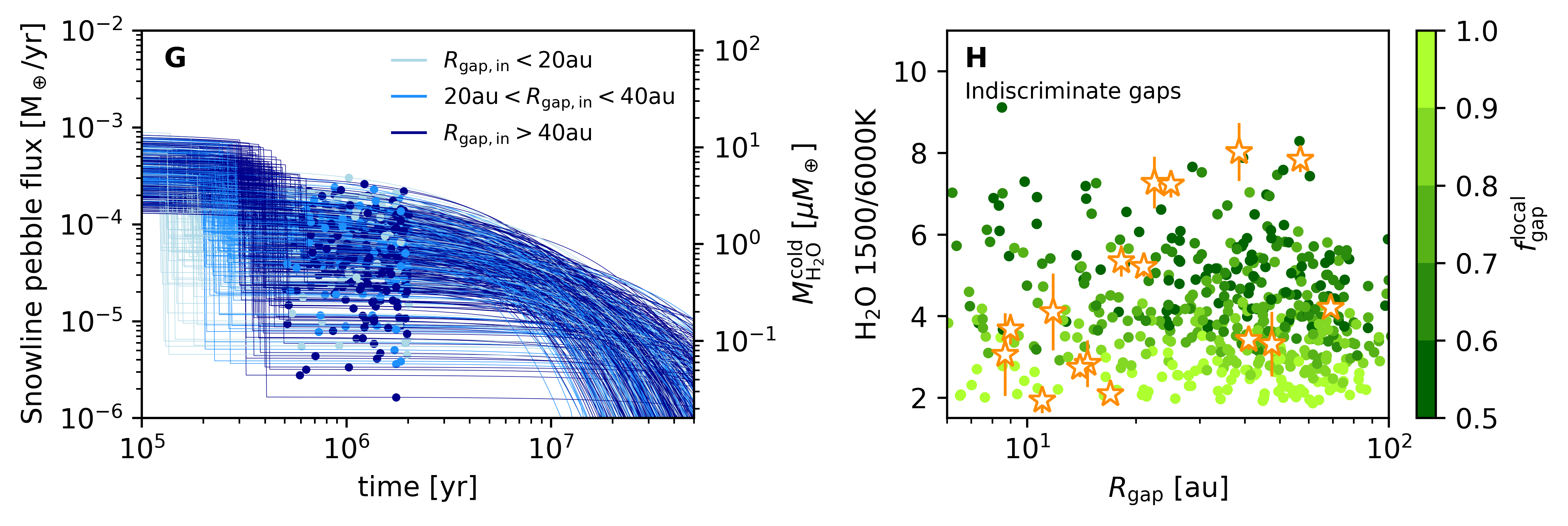}
\caption{Similar to Fig.~\ref{fig:popsynth_1} but for alternative scenarios (see Table~\ref{tab:model_params}) in which: drift is more rapid (top row), gaps are leakier (second row), gaps appear later (third row), or gaps are indiscriminate, i.e., the reduction in snowline pebble flux does \emph{not} depend on the gap's location (bottom row).}\label{fig:popsynth_2}
\end{figure*}

An in-depth exploration of the impact of all parameters involved is beyond the scope of this study but we briefly discuss here the consequences of changing the timing of pebble delivery, the leakiness of gaps, and the timing of gap formation, concepts that have received considerable attention recently and that connect closely to the larger story of planet formation \citep[e.g.,][]{drazkowska_2021, segura-cox2020, stadler2022,ohashi2023, delussu2024, vioque2025}. Specifically, we present in Fig.~\ref{fig:popsynth_2} the results of a `rapid drift' series of models in which pebbles grow larger and drift faster due to a lower turbulence and higher fragmentation velocity (see Table~\ref{tab:model_params}), a `leaky gaps' scenario (in which the $f_\mathrm{gap}^\mathrm{local}$ is lowered to between 0.5 and 0.9), a `later gaps' series (in which $t_\mathrm{gap}=10^6\mathrm{~yr}$), and an `indiscriminate gaps' case (in which the snowline pebble flux reduction is made independent of gap position).

Comparing the data and models in Fig.~\ref{fig:popsynth_2}B it is evident that models in which a significant amount of pebble drift happens early (i.e., before gaps open) are not capable of reproducing the observed trends. This finding seems to match other studies highlighting the need for early substructure to slow down drift in order to reproduce observables such as mm fluxes, disk sizes, and spectral indices \citep[e.g.,][]{pinilla2012,delussu2024}. 

The permeability (i.e., leakiness) of gaps is an open question as various computational studies have highlighted that small grains can pass through planet-induced gaps relatively easily \citep[e.g.,][]{zhu2012, weber2018, drazkowska2019, stammler_2023, vanclepper2025}. Our leaky gap scenario (Fig.~\ref{fig:popsynth_2}D) does reproduce a range of line ratio values but fails to capture the trend with gap position, producing too many high line ratios for $R_\mathrm{gap}\lesssim20\mathrm{~au}$. For this particular sample of disks, we therefore favor relatively effective gaps, at least for the ones inside 20 au. If these gaps themselves turn out not to be capable of acting as effective drift barriers, additional effects such as the formation of planetesimals inside the associated pressure bumps \citep[e.g.,][]{carrera2019,lau2022} and their impact on water enrichment \citep{kalyaan_2023, danti2023} would have to be considered.

The models with later gaps (Fig.~\ref{fig:popsynth_2}F) capture broadly the variation with $R_\mathrm{gap}$ but tend to underpredict line ratios across the board as the late appearance of gaps has allowed more pebbles to drift through during the first Myr. Moreover, some of later gap models have high line ratios because they are captured just between $t_\mathrm{gap}$ and $t_\mathrm{gap}+t_\mathrm{delay}$ (i.e., after the gap formed but before its impact in the disk is felt, see Eq.~\ref{eq:delay}), which is a relatively short-lived phase and one that may be better captured in models with a (more) gradual gap opening treatment.  Based on this exercise we conclude that to reproduce the observed trends of Fig.~\ref{fig:data_trends} our models favor scenarios in which pebble drift is relatively slow and most gaps open within the first Myr. 

As a final test we revisit the main working hypothesis in our models, based primarily on \citet{kalyaan_2021, kalyaan_2023}, that close-in gaps may be the most effective in reducing inner water enrichment through pebble drift. To do so, in the bottom row of Fig.~\ref{fig:popsynth_2} we present the case where gap position does \emph{not} play that role (specifically, we have omitted the exponential term in Eq.~\ref{eq:gap_impact}). Including a spread in $f_\mathrm{gap}^\mathrm{local}$ values (between 0.5 and 0.99) does result in a range of line flux ratios (with higher $f_\mathrm{gap}^\mathrm{local}$ resulting in lower cold water masses, as expected, see Fig.~\ref{fig:popsynth_2}H), but the trend with $R_\mathrm{gap}$ is absent. From this test we conclude that our models can only reproduce the behavior found in Fig.~\ref{fig:data_trends} if their impact is indeed position-dependent.

\section{Discussion and conclusions}\label{sec:discussion}
In this study we have used a specific diagnostic (the 1500/6000 K water line flux ratio from JWST/MIRI) to target a specific reservoir (the observable cold water vapor in the disk surface) to investigate a process in a specific disk region (icy pebbles drifting across the midplane water snowline) to conclude that, in disks where substructure has been resolved, it is the innermost gaps that regulate water enrichment from ice drift into the terrestrial planet formation region (Fig.~\ref{fig:data_trends}). This conclusion, reached for the first time by combining JWST and ALMA data, supports predictions from recent modeling works \citep{kalyaan_2021, kalyaan_2023, easterwood_2024, mah_2024}. The population synthesis exploration performed in this work provides a fiducial model that can reproduce the observed trend in structured disks based on pebble drift models (Fig.~\ref{fig:popsynth_1}), disfavoring scenarios where pebble drift happens early, where gaps are very leaky, or gaps form late (Fig.~\ref{fig:popsynth_2}). 

We remark, however, that the scenarios explored here are not necessarily universal and that water delivery in individual disks may be regulated by a combination of processes. Upcoming JWST programs aimed at expanding the sample of systems observed at different ages, across different spectral types, and across a range of mass accretion rates \citep[e.g.,][]{zhang_jwst, long_jwst, huang_jwst, majo_jwst}, will be fundamental in providing additional multi-dimensional observational constraints. Moreover, increasing the sample of disks with uniform high-resolution ALMA continuum data will further test the correlation described in Sect.~\ref{sec:data}. To confirm the conclusions of this work, deep dust gaps that provide effective traps to ice-delivering pebbles should only be found with low 1500/6000~K water ratios in MIRI spectra, unless other processes affect the emission (e.g. a dust cavity or accretion bursts, see Section \ref{sec:data}). An example might be the case of BP~Tau, where an inner dust cavity has recently been suggested from high-resolution ALMA data \citep{zhang_2023,gasman2025}; if this is confirmed, the 1500/6000~K ratio might be slightly increased by dust dispersal rather than water delivery \citep[][and Mallaney et al. in prep.]{banzatti_2017}.

Building on \citet{romero-mirza_2024,banzatti_2025}, our methodology in this work focused on connecting pebble mass fluxes to cold water masses and line diagnostic ratios (Sect.~\ref{sec:drift_to_cold_water}). While this approach captures the main factors involved in the story, we recognize that the water snowline region is characterized by complex interactions between thermochemistry, hydrodynamics, dust/ice evolution, and planet(esimal) formation \citep[e.g.,][]{hartmann2017, lichtenberg_2021, woitke2022, bosman2022, houge_2025, wang2025}. Additionally, the full story of how the chemical and physical structure of the entire inner regions of protoplanetary disks (their midplanes \emph{and} surface layers) respond to various degrees of pebble drift has to also include evolving elemental ratios (e.g., C/O) and molecular abundances for all relevant and observable species \citep[e.g.,][]{xie2023, sellek2025, long2025}. The full JWST/MIRI spectra contain a wealth of information beyond the line ratio diagnostic used here that may help shed light on the bigger picture \citep[e.g.,][]{gasman2025, arulanantham2025}. 

The remarkable outcome of this work is that, in its simplicity, a population synthesis based on a few processes combined to new sensitive water line ratios from JWST and the highest-resolution ALMA observations finds support to recent models of dust evolution and pebble accretion to place them in context of the emerging exoplanet diversity and Solar System formation (Fig.~\ref{fig:pebble_flux_overview}). Specifically, one aspect that is emerging is that the high pebble fluxes required to form Super-Earths in \citet{lambrechts_2019} (the top of the orange area in Fig.~\ref{fig:pebble_flux_overview}), may be provided primarily by disks with no (deep) gaps in the inner ${\sim}20\mathrm{~au}$. Planets that form in these disks may accrete gas in water-vapor-rich environments, inheriting high O/H and low C/O ratio atmospheres \citep[e.g.,][]{bitsch2021, mah_2024}. Conversely, disks with deeper gaps closer in are characterized by a rapidly decreasing pebble flux, perhaps more in line with what has been inferred for the inner Solar System \citep[Fig.~\ref{fig:pebble_flux_overview} and][]{kruijer_2020, lichtenberg_2021}, suggesting a water-vapor-poor rocky planet formation environment. While bulk Earth is indeed very `dry' \citep{krijt_2023} and the existence of a class of water-rich small planets (at least around M-dwarfs) has been argued to exist \citep{luque_2022}, we have to remember terrestrial planets form primarily from solid building blocks and that water may be lost during the pebble accretion \citep{johansen_2021, wang2023} and planetesimal \citep{lichtenberg_2019, lichtenberg_2023} stages as well. Ultimately the observational trends presented in this study will have to be folded into state-of-the-art planet formation models to assess the impact of substructure and pebble drift (or the lack thereof) on emerging planetary properties.

\begin{acknowledgments}
We thank the reviewer for providing insightful comments that helped improve the manuscript. SK is grateful to Joanna Dr{\k{a}}{\.z}kowska and Tim Lichtenberg for sharing and discussing the Solar System model in Fig.~\ref{fig:pebble_flux_overview}. SK and TK acknowledge support from STFC Grant ST/Y002415/1. AB acknowledges support from JWST-GO-01640. KZ acknowledges support from JWST-GO-01584. This project was partially supported by the STScI grant JWST-GO01584 ``A DSHARP–MIRI Treasury survey of Chemistry in Planet-forming Regions''. PP acknowledges funding from the UK Research and Innovation (UKRI) under the UK government’s Horizon Europe funding guarantee from ERC (under grant agreement No 101076489). Part of this research was carried out at the Jet Propulsion Laboratory, California Institute of Technology, under a contract with the National Aeronautics and Space Administration (80NM0018D0004). This work benefited from information exchange within the program ‘Alien Earths’ (NASA Grant No. 80NSSC21K0593) for NASA’s Nexus for Exoplanet System Science (NExSS) research coordination network. The JWST data presented in this article were obtained from the Mikulski Archive for Space Telescopes (MAST) at the Space Telescope Science Institute. The specific observations analyzed can be accessed via \dataset[doi: 10.17909/cgfc-jx37]{\doi{10.17909/cgfc-jx37}}.

\end{acknowledgments}





%

\facilities{JWST(MIRI), ALMA}

\software{astropy \citep{2013A&A...558A..33A,2018AJ....156..123A,2022ApJ...935..167A}, iSLAT \citep{iSLAT,iSLAT_code,islat_3}
          }


\appendix

\section{Connecting cold water mass to 1500/6000K ratios}\label{sec:appendix_B}

\begin{figure*}[ht!]
\centering
\includegraphics[width=0.8\textwidth]{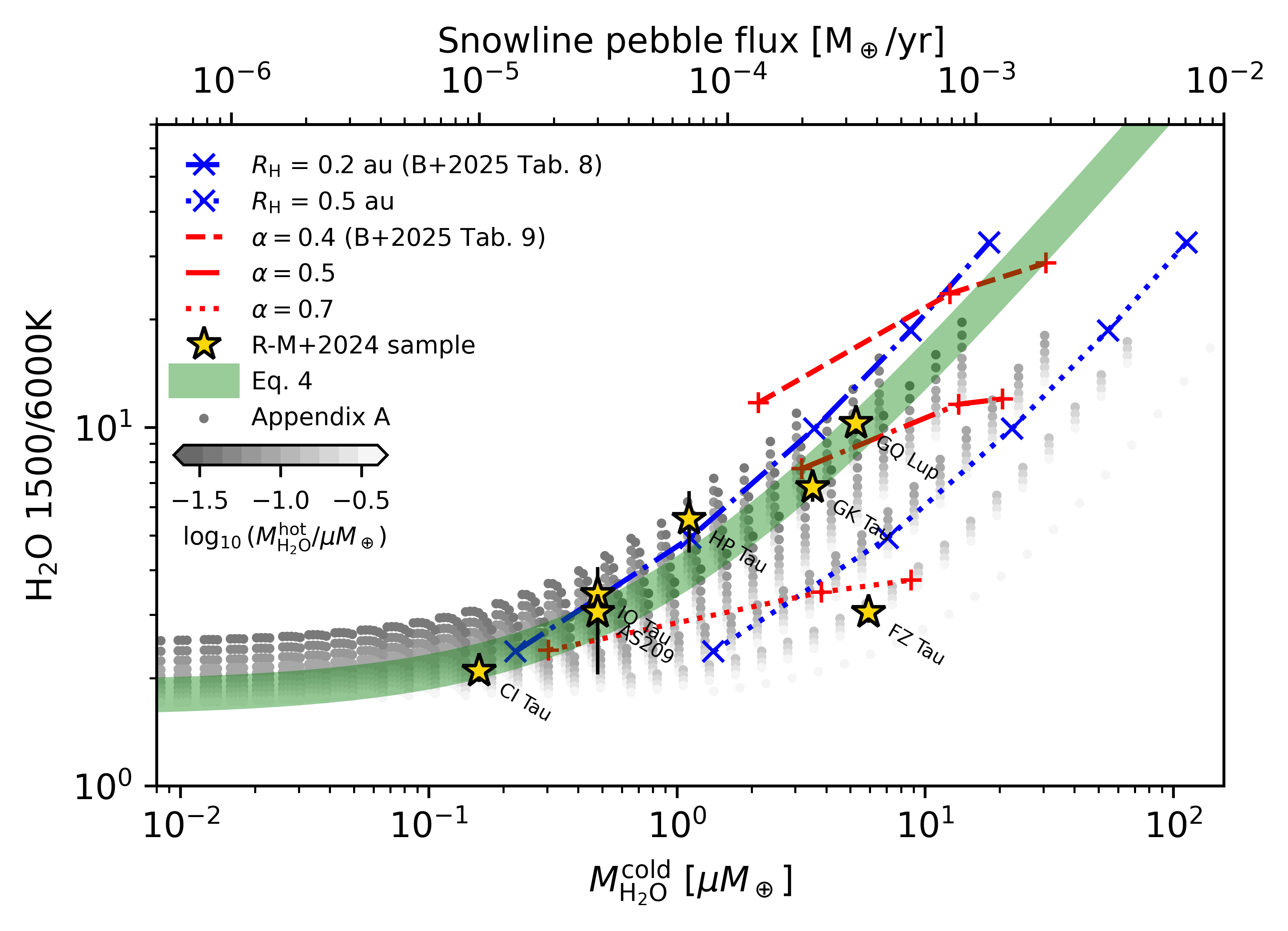}
\caption{Inspired by the model series in \citet{banzatti_2025} (shown in blue and red for different values of the hot water emitting radius $R_\mathrm{H}$ and temperature powerlaw index $\alpha$) and additional multi-component slab models presented in Appendix \ref{sec:appendix_B} (grey points), we employ an empirical relation (Eq.~\ref{eq:h2oratio}, shown in green) to convert cold water masses (bottom x-axis) to 1500/6000K line ratios. Gold stars show $M_\mathrm{H_2O}^\mathrm{cold}$ for seven objects as obtained by \citet{romero-mirza_2024} through fitting the entire rotational part of the water spectrum against line ratios for the same objects as reported in \citet{banzatti_2025}.}
\label{fig:hockey_stick_figure}
\end{figure*}

Figure \ref{fig:hockey_stick_figure} compiles published cold water masses and water line flux ratios from \citet{romero-mirza_2024} and \citet[][Tables 8 and 9]{banzatti_2025}. To explore the connection between these quantities further and to extend the range of cold water masses covered, we set up an additional grid of slab models with three components (hot: $850\, \mathrm{K}$, warm: $400\, \mathrm{K}$, and cold: $190\, \mathrm{K}$) using DuCKLinG \citep{Kaeufer2024} (shown in grey in Fig.~\ref{fig:hockey_stick_figure}). DuCKLinG interpolates and integrates a grid of 0D LTE slab models \citep{Arabhavi2024} with water line data taken from HITRAN2020 \citep{Gordon2022} to derive the fluxes in the MIRI wavelength range. For this exercise, the emitting areas ($A_i$) of all slab components are quantified using the emitting radius ($R_i$) using $A_i=\pi R_i^2$. The radius of the warm component is double the radius of the hot component ($R_\mathrm{W}=2R_\mathrm{H}$), with the radius ($R_\mathrm{C}$) of the cold component being varied between $R_\mathrm{W}$ and $10R_\mathrm{W}$ on a logarithmic scale. The column density of the warm component is fixed to $5\times 10^{17}\,\mathrm{cm^{-2}}$ and the column densities of the hot and cold component are varied between $10^{18}-10^{19}\,\mathrm{cm^{-2}}$ and $10^{16}-10^{19}\,\mathrm{cm^{-2}}$, respectively\footnote{Additionally, we restrict the grid to models with column densities of the cold component smaller than the column density of the hot component. This avoids unrealistic parameter combinations that contradict typical output from thermochemical models \citep[e.g.][, M.~Vlasblom et al. submitted]{Kaeufer2024}.}. All temperatures, column densities, and radii are inspired by Table~8 from \cite{banzatti_2025} who set up slab models to overlap with the observed water line ratios. The cold water mass of every model is calculated as the mass of the cold slab component. The fluxes of a set of unblended water lines \citep[Table 1 from][]{banzatti_2025} are integrated. The line tracing the hot water component has an upper level energy of $6052\,\mathrm{K}$, while the two lines whose sum is used to trace cold water have upper level energies of $1448\,\mathrm{K}$ and $1615\,\mathrm{K}$, respectively. The ratio of these integrated fluxes compared to the models' cold water mass is shown in Fig.~\ref{fig:hockey_stick_figure}. 

For cold water masses exceeding ${\sim}0.1-1 \mu M_\oplus$, Fig.~\ref{fig:hockey_stick_figure} shows the 1500/6000K line flux ratio is sensitive to the amount of cold water -- and hence pebble flux (top x-axis). This behavior is seen across the slab models, the model series from \citet{banzatti_2025}, and the fits of \citet{romero-mirza_2024} to the full rotational rotational part of the spectrum. For the lowest cold water masses, relatively constant line ratios are predicted. This can be understood when considering that small cold water masses result in low fluxes of the examined cold water lines, which will make the typically small contribution from hotter water dominate also this wavelength region and therefore the ratio, i.e., the influence of changes from the cold component is minimized. For example, the integrated flux of the cold water lines for a cold water slab model with a mass of ${\sim}0.88\,\mathrm{\mu M_\oplus}$ (temperature: $400\,\rm K$, column density: $10^{18}\,\mathrm{cm^{-2}}$, and emitting radius: $0.5\,\mathrm{au}$) is about $7.5$ times larger than the flux of a typical hot water component (temperature: $850\,\rm K$, column density: $10^{18}\,\mathrm{cm^{-2}}$, and emitting radius: $0.1\,\mathrm{au}$). However, for a model (temperature: $400\,\rm K$, column density: $10^{18}\,\mathrm{cm^{-2}}$, and emitting radius: $0.2\,\mathrm{au}$) with a very low water masses ${\sim}0.14\,\mathrm{\mu M_\oplus}$ the integrated flux is only ${\sim}1.2$ the integrated flux of the hot component.

Apart from the cold water mass, the 1500/6000K ratio also depends on the remaining water reservoir. The gray points in Fig.~\ref{fig:hockey_stick_figure} show how variations in $M_{\ce{H_2O}}^\mathrm{hot}$ of about an order of magnitude (similar to the spread observed in \citealt{romero-mirza_2024}) introduce a scatter for identical $M_{\ce{H2O}}^\mathrm{cold}$. The larger amount of hot water in FZ Tau \citep[][Fig.~9]{romero-mirza_2024} may thus be responsible for the apparent offset of that system in Fig.~\ref{fig:hockey_stick_figure}, although we note that FZ Tau is unusual also due to its `ring-like' water vapor distribution and high mass accretion rate \citep[][Sect.~4.3]{romero-mirza_2024}. For the purposes of this study we adopt the trend described by Eq.~\ref{eq:h2oratio} (shown in green in Fig.~\ref{fig:hockey_stick_figure}) that appears broadly consistent with existing water analyses \citep{romero-mirza_2024}, but caution that variations in the amount and distribution of hot water may introduce significant scatter. These variations likely reflect differences in disk temperature and structure, but a connection to pebble drift cannot be fully excluded.

\section{Additional plots for the sample}\label{sec:names}

\begin{figure*}[h]
\centering
\includegraphics[width=0.9\textwidth]{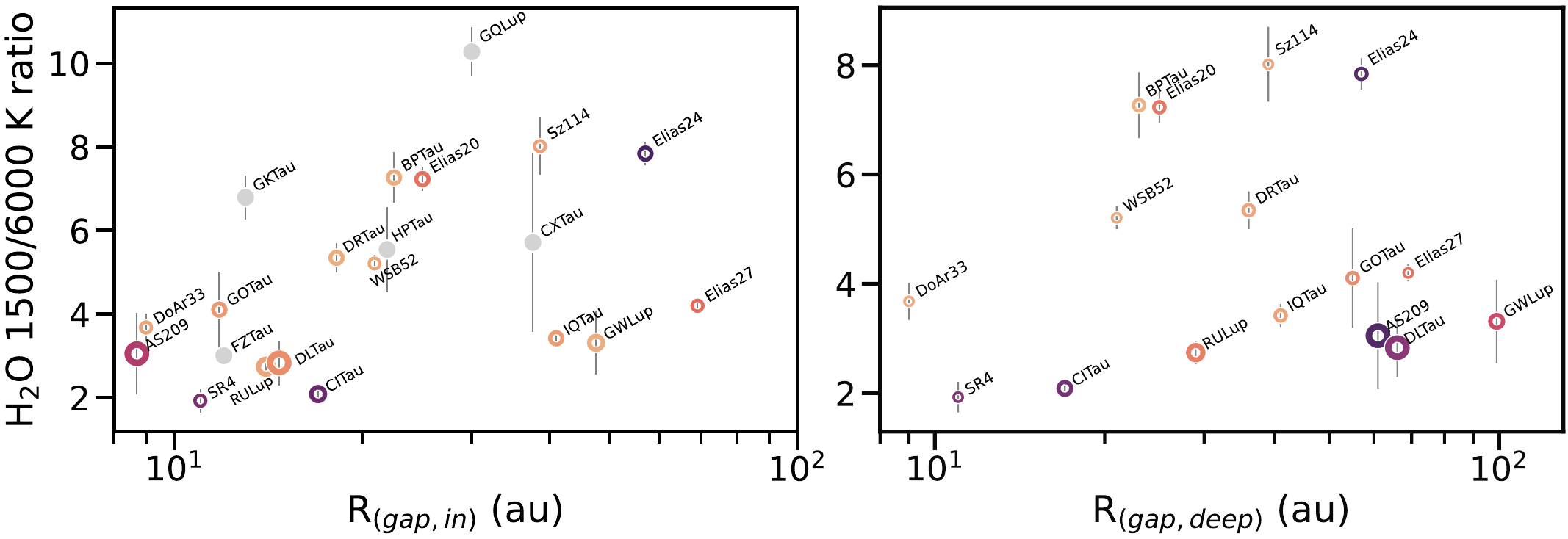}
\caption{Left: same as right panel in Figure \ref{fig:data_trends} but including target names for reference. Right: using the radial location of the deepest gap in each disk rather than the innermost gap.}
\label{fig: data_figure_names}
\end{figure*}

Figure \ref{fig: data_figure_names} includes target names for the right panel in Figure \ref{fig:data_trends}, for guidance on identifying individual disks.
The right panel in the figure shows the scatter of points when using the location of the deepest gap, taken from the same references as shown in Table \ref{tab: sample}, rather than the innermost gap in each disk; the Pearson correlation decreases to -0.05 with p-value of 0.86.


\bibliography{water_substructure}{}
\bibliographystyle{aasjournalv7}



\end{document}